%% file: berggren-lcws10-np-proceedings.tex
\documentclass[twoside]{ilcws10}
\usepackage[latin1]{inputenc}
\usepackage[dvips]{graphicx,epsfig,color}
\usepackage{wrapfig,rotating}
\usepackage{amssymb,amsmath,array}
\usepackage{mdwlist}
\usepackage[pdfborder={0 0 0 0}]{hyperref}
\usepackage{pgfpages}

\input{newcom}

\begin{document}
\title
{ILC Beam-Parameters and New Physics}
%***********************************************************************
% AUTHORS INFORMATION AREA
%***********************************************************************
\author{Mikael Berggren
% Optional short acknowledgment: remove next line if non-needed
\thanks{Work supported by the DFG through 
the SFB (grant SFB 676/1-2006) and 
the Emmy-Noether program (grant LI-1560/1-1)}
% DO NOT MODIFY THE FOLLOWING '\vspace' ARGUMENT
\vspace{.3cm}\\
% Addresses and institutions (remove "1- " in case of a single institution)
 DESY  \\
Nothkestrasse 85, Hamburg - Germany
}
%%***********************************************************************
% END OF AUTHORS INFORMATION AREA
%***********************************************************************

\maketitle

\begin{abstract}
A brief overview of the linear collider design is given,
with emphasis on the elements of particular importance
for the performance. The modifications of
the RDR design suggested in the SB2009 proposal are
presented, once again with emphasis on those item that
have most impact on the performance.
In particular, the effects on New Physics channels
are studied, by two examples:
the analysis of the properties of $\stau$:s in the
SUSY benchmark point SPS1a', and the model-independent
Higgs recoil mass analysis.
It is shown that for both these cases, the SB2009
design performs significantly worse than the RDR design:
For the \stau ~analysis, the uncertainties on both
the mass and cross-section determination increases
by 20 \% (or 35 \% if the travelling focus concept
is not deployed). For the Higgs analysis,
the corresponding increase in uncertainty is found 
to be 70 \% both for cross-section and mass (or 100 \%
without travelling focus).
For both channels, the deterioration is to a large
part due to the move of the positron source to the
end of the linac.
\end{abstract}

\section{Introduction}
This note is a combination of two talks given at LCWS 2010.
One talk was given in the machine-detector interface
session, and was mainly aimed in explaining our findings
on the physics impact of different options for the
design of the ILC to the accelerator community. 
The
other invited talk had the opposite focus: to explain
to the physics community 
what options were discussed for the machine and
how and why they influence the physics reach of the ILC.
The latter talk was given in the SUSY and New Physics
session.

The note is organised as follows: In the first section,
a brief description of the different subsystems of the ILC
is given, with particular emphasis on which design choices
mostly influence the performance, in terms of total
luminosity, luminosity on-peak, polarisation, 
quality of the beams, machine background, and energy reach.
The energy dependence of these factors is also discussed.
The second section presents the SB2009 proposal,
and in the third section the impact of the SB2009 
proposal on the performance  
on two
key-channels (\stau:s in the SUSY point SPA1a' and
the SM Higgs) is discussed.
Finally, conclusions are given, and a brief summary 
is done of
the current activities within the GDE to alleviate the
performance problems arising from the new base-line
design pointed out here and in other contributions
to LCWS 2010.
\section{The linear collider}

In an ideal $e^+e^-$-collider, one would have
an exactly known initial $e^+e^-$ state.
Both beams would be fully polarised,
and the intensity of the beams would be such
than one would have as many events as needed not to
be statistics-limited at 
any $E_{CMS}$ for all interesting physics channels.
Furthermore, one would have pure electron/positron beams,
with negligible background from the machine or from
$\gamma\gamma$ events.

Clearly, none of these properties will be present in any
actually buildable accelerator. On the contrary,
in a real linear collider, the
beam energy has both initial and beam-beam induced spread.
The degree of positron polarisation will be rather low ($\sim$ 30 \%), 
and electron polarisation will be $<$ 100 \%.
Obviously, the luminosity will be limited, so that many
physics observables will remain statistics limited, even at
the end of the life-span of the accelerator.
Due to beam-beam effects, the beams will be mixed lepton and photon ones,
and there will be huge numbers of low energy background 
particles from the machine. Clearly, there is no way to
avoid background from $\gamma\gamma$ events.

\begin{wrapfigure}{r}{0.5\columnwidth}
%\begin{figure}
\centerline{\includegraphics[scale=0.38]{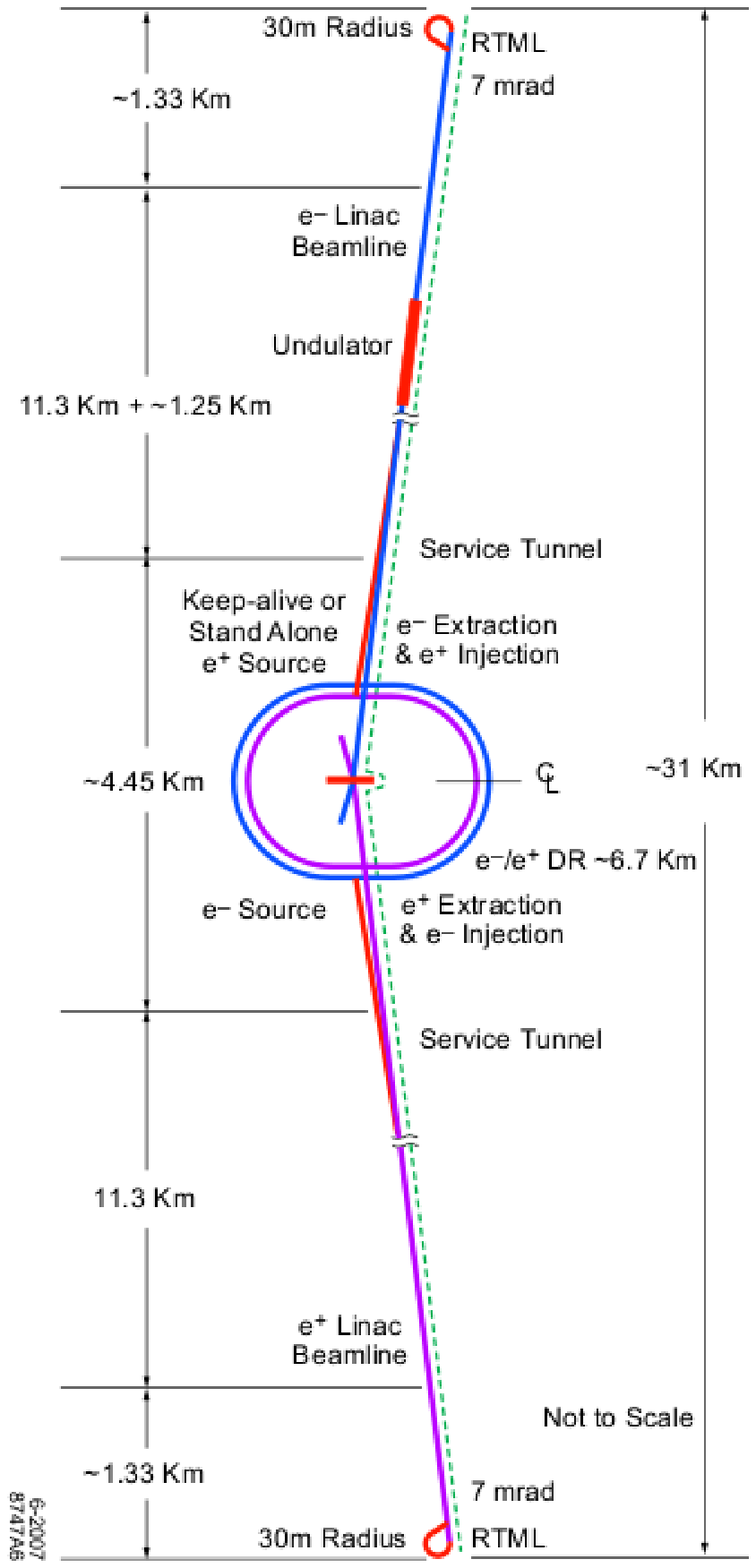}
\includegraphics[scale=0.45]{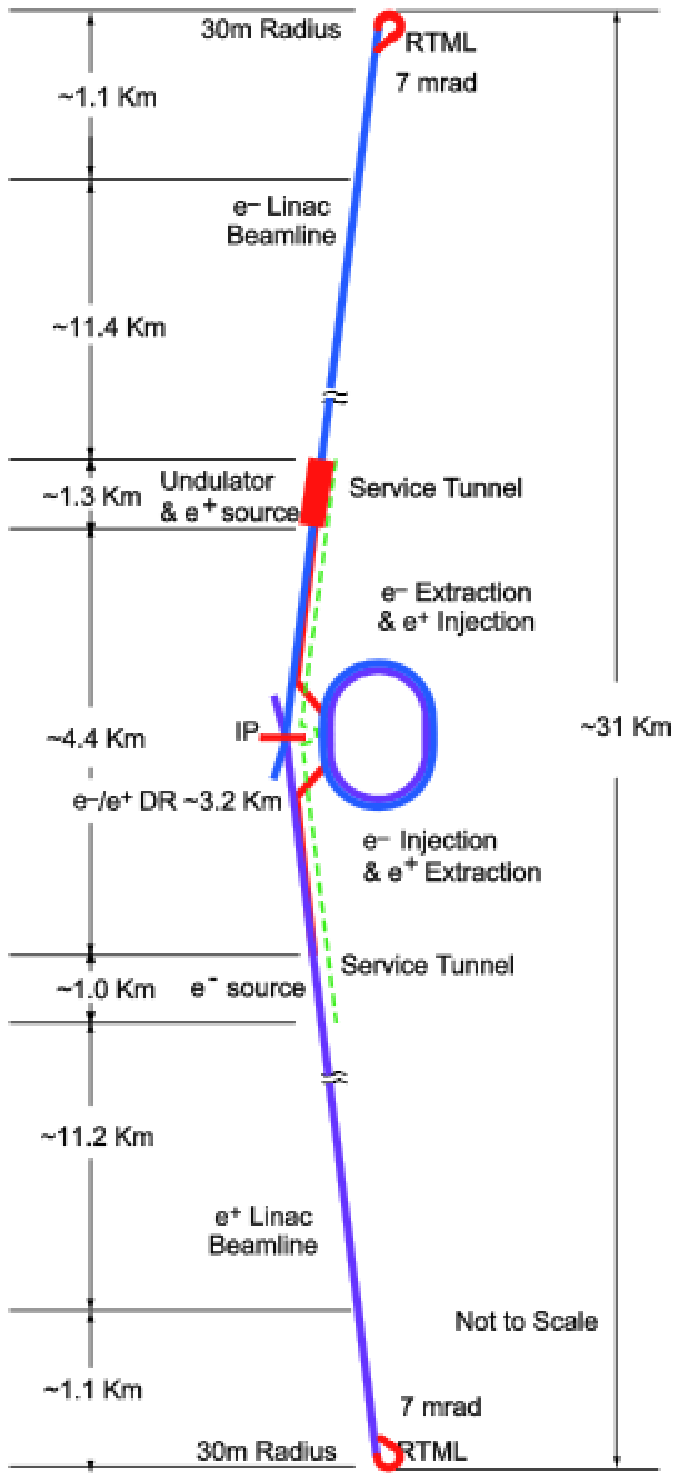}}
\caption{ILC, RDR~\cite{RDR} (left) and SB2009~\cite{SB2009} (right) designs }\label{Fig:ILC}
\end{wrapfigure}
%\end{figure}
To create the beams, one must have sources of  electrons and positrons,
and the two need to employ different technologies.
The beams should be as well defined as possible, so the initial random
spread in energy and direction should be reduced, which is done in
the damping system.
The main system of the accelerator, which will allow to
attain the high centre-of-mass energies required is the
main linac.
The beams need to be brought into collision at the centre
of the detectors, and it is the
beam delivery and
final focus systems that are employed to achieve this.

In the following sub-sections, some details are given 
on the implementation of each of these sub-systems at the ILC.
The information is taken from the Reference Design Report~\cite{RDR} (the
RDR), and
Figure~\ref{Fig:ILC} shows schematic layouts for the
RDR and SB2009~\cite{SB2009} designs.

\subsection{Sources}

The electron source comprises
polarised lasers shining on photo-cathodes,
which are specially designed GaAs/GaAsP super-lattice structures
yielding electrons with high polarisation, see Figure~\ref{Fig:eandpsource}.
The emitted electrons are collected and pre-accelerated, 
and then sent to the damping system.
Positrons are obtained by letting
a high  energy electron beam pass
an {helical undulator acting as a FEL, to produce photons of high intensity,
high polarisation and high energy ($\sim 10 \MeV$). 
These hit a 
target to produce $e^+ e^-$-pairs, see Figure~\ref{Fig:eandpsource}.
To avoid damage to the target, it is designed as a rotating wheel.
The electron beam must have an energy of at least $150 ~\GeV$: at lower energies
the positron yield becomes so low that the positron bunches are not filled
to full capacity, and the over-all
luminosity will decrease.

Like the electrons,
the positrons are then collected,
pre-accelerated and sent to the damping system.
The electrons used to produce the photons are the same as those that
will be delivered to the detectors.
Due to the effect of the helical undulator, the electron beam
will therefore obtain an additional energy dispersion.

\subsection{Damping system}
%\begin{wrapfigure}{r}{1.1\columnwidth}
\begin{figure}
\centerline{\includegraphics[scale=0.45]{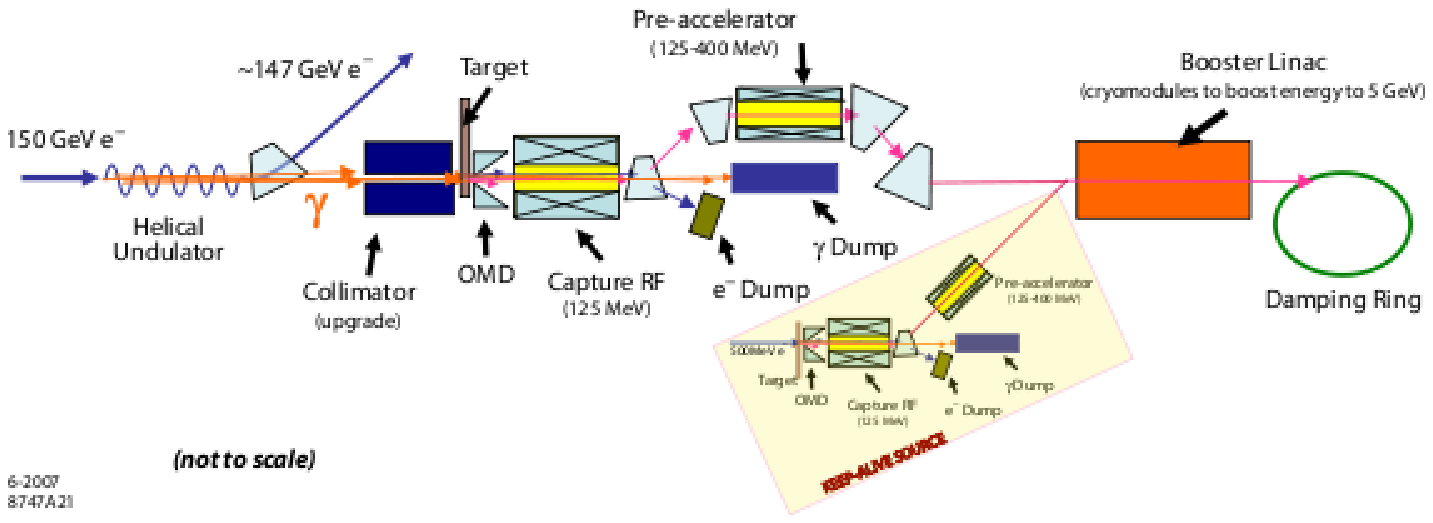}
\includegraphics[scale=0.45]{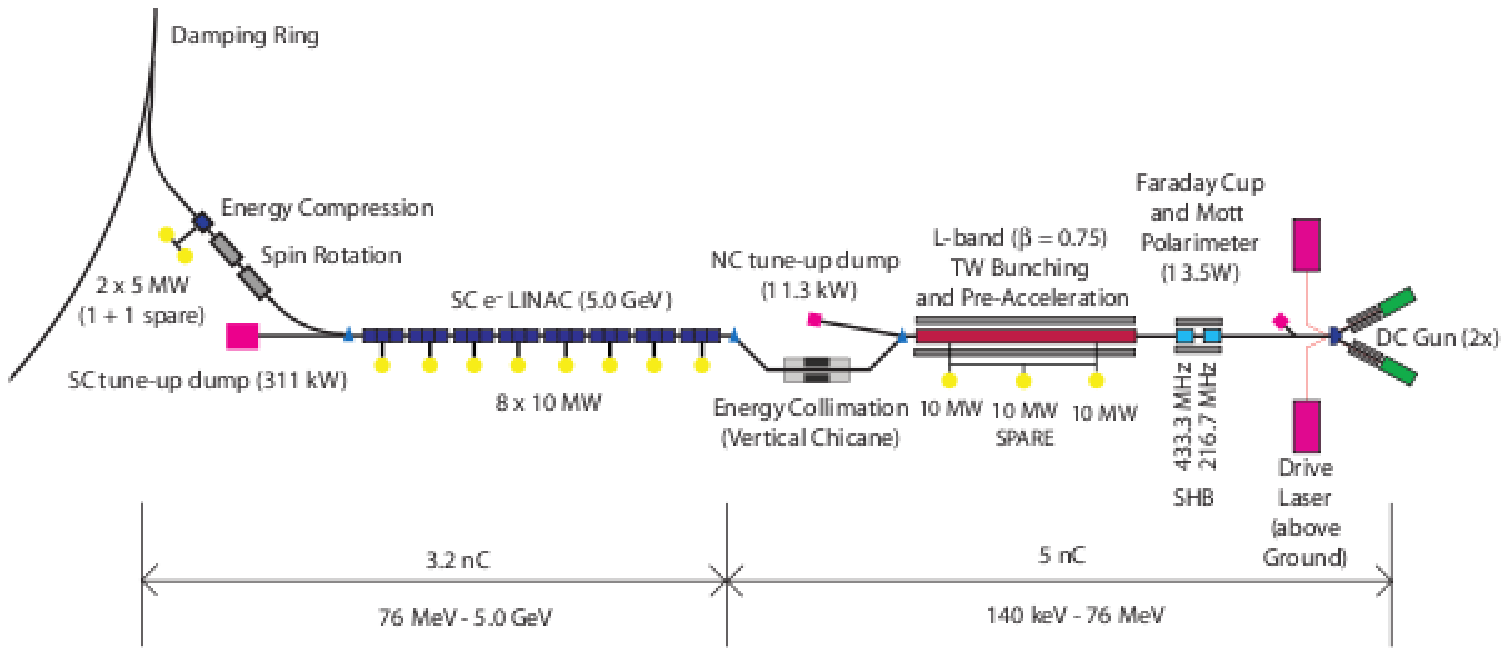}}
\caption{Electron (left) and positron (right) sources. From~\cite{RDR}.}
\label{Fig:eandpsource}
%\end{wrapfigure}
\end{figure}
When emerging from the sources, the dispersion both in angle and
energy of both beams are far to large to be accelerated and delivered
to the detectors.
This dispersion is reduced by
sending the beams (now at $\sim$ 5 \GeV) to rings where they pass
wigglers which make them cool off by synchrotron radiation.
The particle bunches in the damping ring are kicked out,  one-by-one,
every
$\sim$ 100 ns to make a bunch train comprising 
$\mathcal{O}(1000)$ bunches (2625 in the RDR design).
The bunches in the damping-rings are separated by a few ns, given
by the ratio of the circumference of the damping ring to the number of bunches.
Therefore the kickers must be able to switch on or off in a few ns, which
is at the limit of current technology.
This procedure (cooling and bunch-train assembly) must be completed 
in the 200 ms between bunch trains.
The damping rings are at the centre of the complex, so it is needed
to transport the trains $\sim$ 15 km to the start of the
main linac after damping.

\subsection{Main linac}
\begin{wrapfigure}{r}{0.5\columnwidth}
%\begin{figure}
\centerline{\includegraphics[scale=0.45]{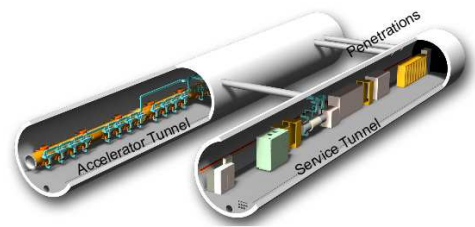}}
\caption{A section of the main linac. From~\cite{RDR}.}\label{Fig:mainlinac}
\end{wrapfigure}
%\end{figure}
The main linac is made of supra-conducting RF cavities each containing 
9 accelerating
cells.
The accelerating gradient is 31.5 MV/m,
which is the foreseeable limit of this technology.
One RF unit contains three cryo-units, two of which contain 9 cavities, 
and one which contains 8
cavities and a focusing quadrupole.
There are 278 RF units in the positron linac, and 282 in the electron one.
There are more in the latter, because the energy lost in the undulator
must be compensated for.
How many particles one can get at the experiment per 
time-unit depends on how much power is fed into the cavities,
which in turn depends on how many klystrons are installed
along the accelerator. How these system are arranged in the RDR 
two-tunnel design
can be seen in Figure~\ref{Fig:mainlinac}.
In the RDR design, the positron source is placed at the point
where the electron beam reaches 150 ~\GeV, about 6 km from the
IP, and  electrons can be delivered 
to the experiment at any energy between 50 and 250 ~\GeV,
by  appropriately accelerating or decelerating them in the
remaining
part of the main linac.

\subsection{BDS and final focus}
%\begin{wrapfigure}{r}{0.5\columnwidth}
\begin{figure}
\centerline{\includegraphics[scale=0.4]{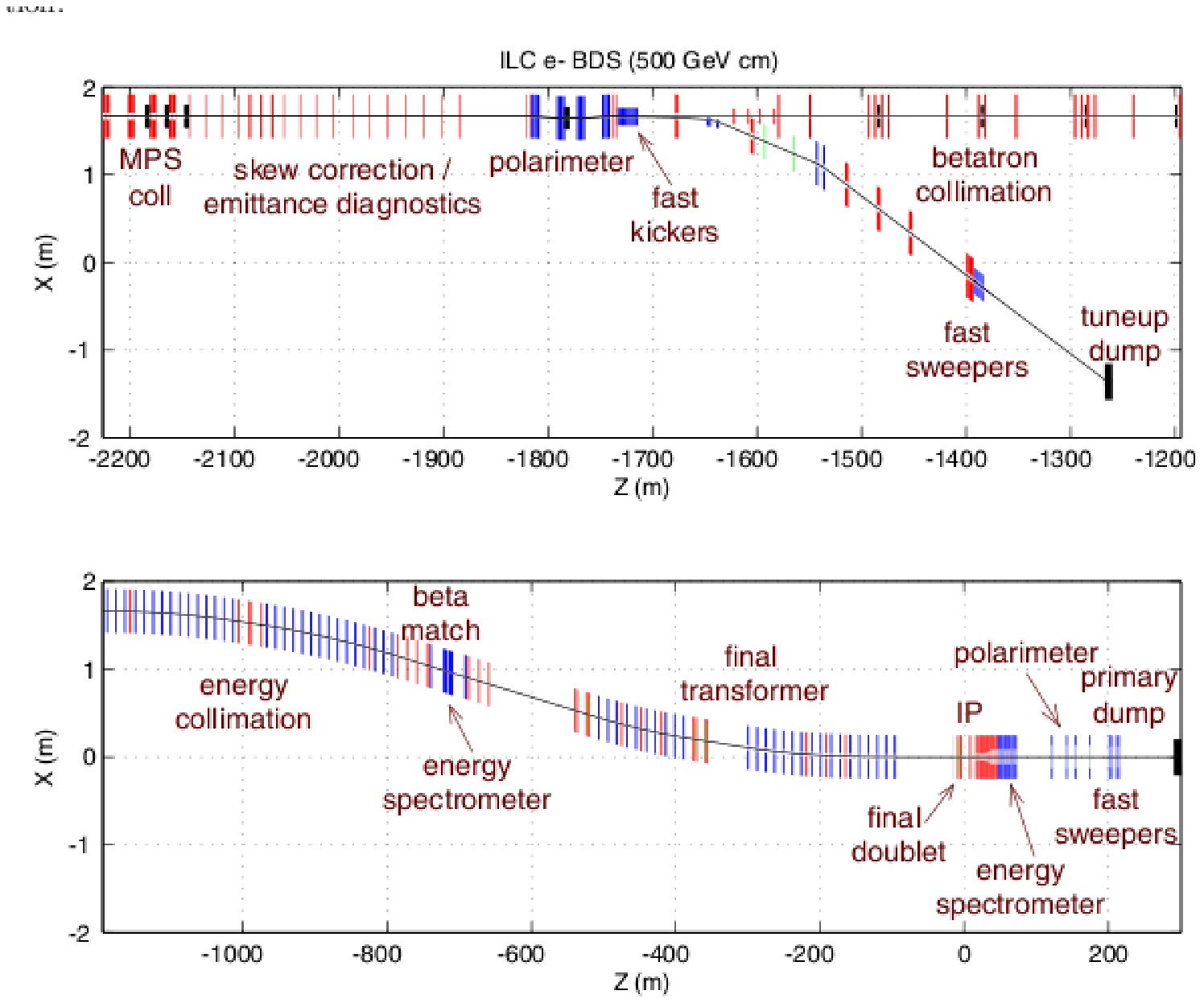}
\includegraphics[scale=0.4]{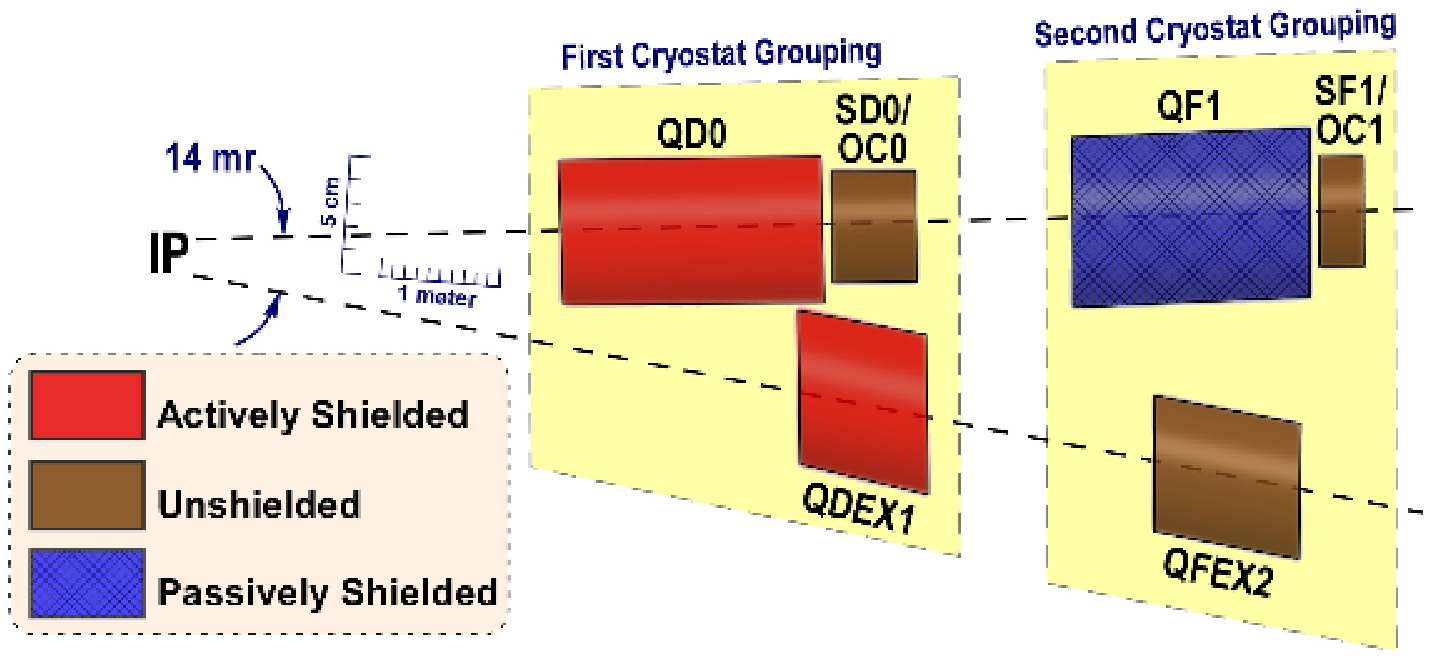}}
\caption{Beam delivery system (left) and Final focus (right). From~\cite{RDR}.}
\label{Fig:BDSandfinalfocus}
%\end{wrapfigure}
\end{figure}
The last two km of the accelerator is the Beam Delivery System (the BDS).
It's purpose is multiple. It should monitor and measure the beam-properties, 
and clean the beam from halo particles.
It should also protect the detectors,
since anything the beam hits at this location will give secondaries
- E$_{beam}$
could be up to 500 \GeV~ - that might hit and
damage the detectors. A schematic of the BDS is shown to the left in 
Figure~\ref{Fig:BDSandfinalfocus}.

\begin{wrapfigure}{r}{0.5\columnwidth}
%\begin{figure}
\centerline{\includegraphics[scale=0.28]{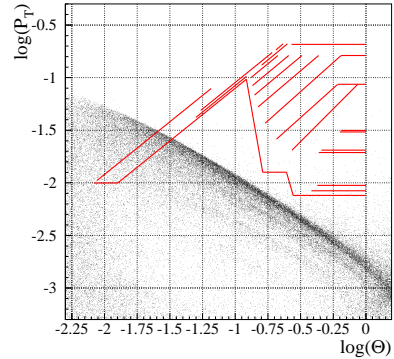}}
\caption{Beam-strahlung particles in ILD.}\label{Fig:pairsrdrnom}
\end{wrapfigure}
%\end{figure}
The final focus system (Figure~\ref{Fig:BDSandfinalfocus}, right) is 
the last 20 m before the detectors. It
focuses the beams to few 100 nm horizontally, and 
few nm vertically at the centre of the detector.
There are several limiting factors on how strongly the beams
can be focused, both from fundamental optics, from beam-beam
interactions and from stability against ground-motion.
In addition, stronger focusing inevitably induces more
background in the detectors.

\subsection{Beam-strahlung}

 Due to the very strongly focused beams, both the electric
and magnetic fields induced by one beam
have a large bending power on the other one.
As a consequence
the primary beam is focused by the other beam,
and will emit a large amount of 
synchrotron radiation~\cite{beambeam}\cite{schulte}. 
The emission of synchrotron
radiation widens the distribution of the
primary $e^{\pm}$ energy.
In addition, the synchrotron photons 
can undergo Compton back-scattering, which in turn
has several consequences:
It yields a photon component of beam, 
and a long tail to lower
 energies for the  $e^{\pm}$ beam.
The resulting relative energy-loss due to beam-strahlung is given by
$\delta_{BS} \propto (E_{cm}/\sigma_z)\times(n^2/(\sigma_x + \sigma_y)^2)$,
where $\sigma_{x,y,z}$ are the sizes of the beam in the x,y or z direction,
and $n$ is the number of particles in each bunch~\cite{beambeam}. 
Furthermore, the synchrotron photons can interact with photons (synchrotron ones,
 or virtual ones) in the other beam,  yielding  $e^{\pm}$-pairs.
Necessarily, one of the particles in the
pair will have a charge 
 {\it opposite} to that of its parent beam.
These particles gets {\it de-focused} rather than focused 
by the other beam, and is the origin of the pair background. 
Of these wrong-sign $e^{\pm}$:s, the ones at the outer edge of the beam
will encounter the largest bending force.
The force is independent of the (longitudinal) momentum
of the particle, which means that
$p_{T}$ and $\theta$ anti-correlates for the pair background, 
and that it accumulates at a quite sharp edge in the
$p_{T}$-$\theta$ plane.
To study the effect of the pair background on the detector, 
it is useful to also draw the detector in these
coordinates: 
 Place each detector element at  the $p_{T}$-$\theta$ 
 corresponding to the $p_{T}$ and $\theta$ a particle should 
 have to turn back at 
 the radius and z of the element.
As an example, Figure~\ref{Fig:pairsrdrnom} shows the distribution
of the beam-strahlung pairs in ILD in these coordinates, 
for RDR design with nominal 
beam-parameters. The pairs were generated
with GuineaPig~\cite{schulte}, and the simulation shows 
that there are some 124000 particles 
created per bunch-crossing.
\subsection{Luminosity}
The Luminosity($L$)  is defined as the
density of particles that pass each other per time-unit,
ie. $L=N^2/(t \times A)$, where $A$ is the transverse area of the
beam at the interaction point.
The number of interactions per time unit is therefore $L$ multiplied by
the cross-section for the considered physics process.
For the number of particles passing each other ($N$) per time-unit, one has that
$N^2/t = n^2 N_{bunch} f_{rep}$, where $n$ is the number of particles in the bunch,
$ N_{bunch}$ is the number of bunches 
in the train, and $f_{rep}$ is the number of trains per second.
One can note that RF-power ($P_{RF}$) needed also depends
on $n, N_{bunch},$ and $f_{rep}$ :  
$ P_{RF}= E_{cm} (n N_{bunch} f_{rep}) \times \eta $, where $\eta$ is
the efficiency of the transfer from the RF-system to the beam.
Hence, $L \propto P_{RF}n/AE_{cm}$. 
Furthermore, for $A$, the cross-section of beams at the IP,
one has that
$A \propto \sigma_x \times \sigma_y$, where $\sigma_x$ and $\sigma_y$
are the horizontal and vertical sizes of the beam, respectively.
These sizes are given by the final focus system, the damping system
and the $\gamma$ factor of the beams (proportional to the beam energy):
$\sigma \propto \sqrt{\epsilon \beta} = 
\sqrt{\epsilon_{norm} \beta / \gamma} $.
Here
$\epsilon$ is the emittance, $\epsilon_{norm}$ is the normalised 
emittance - which
is the figure of merit of  the
damping system - and $\beta$, the focusing-power of the final focus system.
It therefore follows that to get high $L$, 
$\sigma_x \times \sigma_y$ should be small. 
However, as mentioned above, the relative energy-loss due to beam-strahlung 
is inversely proportional to  $\sigma_x + \sigma_y$,
so to reduce it, this sum should be large.
The way to achieve small $\sigma_x \times \sigma_y$ and
large $\sigma_x + \sigma_y$ at the same time is to make
the  $\sigma_x$ and $ \sigma_y$ as different as possible, 
ie. to have a flat beam.
If $\sigma_y << \sigma_x$, one has
$\delta_{BS} \propto (E_{cm}/\sigma_z)\times(n^2/\sigma_x^2)$
and by re-ordering of terms that:
$ n/\sigma_x \propto \sqrt{\delta_{BS}\sigma_z/E_{cm}} $.
Therefore one can write a number of scaling-laws for
the luminosity, that puts emphasis on different design-parameters:
\begin{enumerate*}
\item $
L \propto P_{RF} \times \sqrt{\delta_{BS}\sigma_z}/ (\sigma_y E^{3/2}_{cm})
$, which emphasises the dependence on available RF power.
\item $
L \propto  (n N_{bunch} f_{rep}) \times \sqrt{\delta_{BS}\sigma_z}/ (\sigma_y E^{1/2}_{cm})
$, which emphasises the dependence on beam-structure.
\item $
L \propto  (n^2 N_{bunch} f_{rep}) / (\sigma_x\sigma_y) \propto  
(n^2 N_{bunch} f_{rep} E_{cm}) / (\epsilon_{norm} \beta ) 
$,  which emphasises the dependence on beam energy and emittance. Note that the
price for luminosity in $\delta_{BS}$ is hidden in this formulation.
\end{enumerate*}

\section{RDR and SB2009}
The aim of SB2009 proposal~\cite{SB2009} is to save cost, while 
still full-filling the ILC scope document~\cite{scope}.

The main changes wrt. the RDR design is to house the
main linac and it's support systems in a single tunnel 
(meaning a re-design of the RF-system),
half the beam power (meaning half as many 
bunches in the train, which apart from reducing the
need for klystrons and cooling by two, would allow to make the
damping rings smaller),
while keeping the same total luminosity by stronger final focusing, and
to move the positron source to end of the linac 
(easier logistics and higher positron yield).

\begin{wrapfigure}{r}{0.5\columnwidth}
%\begin{figure}
\centerline{\includegraphics[scale=0.29]{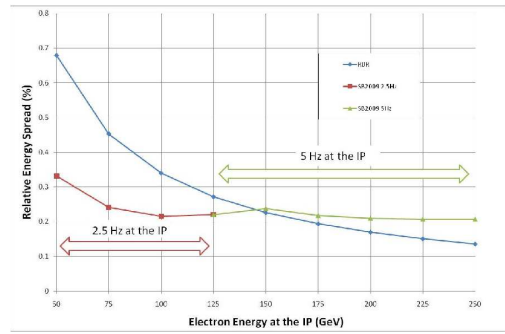}}
\caption{Energy-spread of incoming electron beam. Blue: RDR, 
Green: SB2009 with 5 Hz rep-rate, Red: SB2009 with 2.5 Hz rep-rate. 
From~\cite{SB2009}.}\label{Fig:Espread}
\end{wrapfigure}
%\end{figure}
For the performance of the machine,
this design has a number of issues:
The decrease of the beam-size will increase $\delta_{BS}$,
meaning that the luminosity within 1 \% of nominal 
reduced from 0.83 to 0.72 at 500 \GeV.
It also gives more overlaid tracks in the detector,
and twice as much energy in the low angle calorimeter
(the BeamCal).
The halving of the number of
bunches at constant total luminosity 
will double the luminosity per bunch-crossing,
and hence the probability to have a $\gamma\gamma$ event
in the same bunch-crossing as a physics event.
The move of the positron source will give lower
luminosity below 300 \GeV, since - as described above -
once the electron beam is below 150 \GeV, the positron
bunches will not be full. Since the source is
at the end of the linac, the option to decelerate the beam after
the undulator no longer exists. At 250 \GeV, the luminosity
would only be third of the RDR value.
In addition, at 500 \GeV, the electron beam energy-spread
will
increase from  0.16 \% to 0.21 \% and the positron polarisation 
will 
decrease from 33 \% to 22\%, as can be seen in
Figures~\ref{Fig:Espread} and~\ref{Fig:polandyield}.

\begin{wrapfigure}{r}{0.5\columnwidth}
%\begin{figure}
\centerline{\includegraphics[scale=0.5]{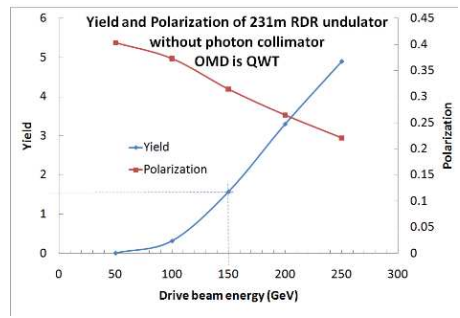}}
\caption{Positron yield and polarisation. From~\cite{SB2009}.}\label{Fig:polandyield}
\end{wrapfigure}
%\end{figure}
It should be noted that the total luminosity 
would be unchanged wrt. RDR only if a novel focusing
scheme - the Travelling Focus (TF) scheme~\cite{TF}- can be made
practical. If not, the SB2009 proposal would yield
a total luminosity 25 \% lower than the RDR design
already at 500 \GeV.

As mentioned above, the increase of  $\delta_{BS}$
increases the number of beam-strahlung pairs per bunch-crossing
by
approximately a factor two. The detector-elements for which this
is most likely to pose a problem is the vertex-detector,
and the BeamCal. In ILD, the vertex detector integrates
over a fixed time-window, which is much longer than the
intra-bunch separation, so given the fact that
the number of bunches per train is reduced by a factor
two in the SB2009 design, the number of hits in each
time window will be approximately the same for the two
designs~\cite{SB2009vtx}. 
The BeamCal, on the other hand, has a read-out fast enough
to read single bunch-crossings.
As can be seen in Figure~\ref{Fig:bcalnumberanddensity}, showing hit
and energy densities for the two designs, the SB2009 would give
a doubling of the background levels.
The plots show only the GuineaPig simulation, not
the full detector simulation, but does include effects of
the crossing-angle and compensating magnetic field (the anti-DID  field).
The issue is whether  one still can detect a $\approx$ 250 GeV
       electron from a $\gamma\gamma$ process over the increased
pair-background.

\begin{figure}
\centerline{\includegraphics[scale=0.32]{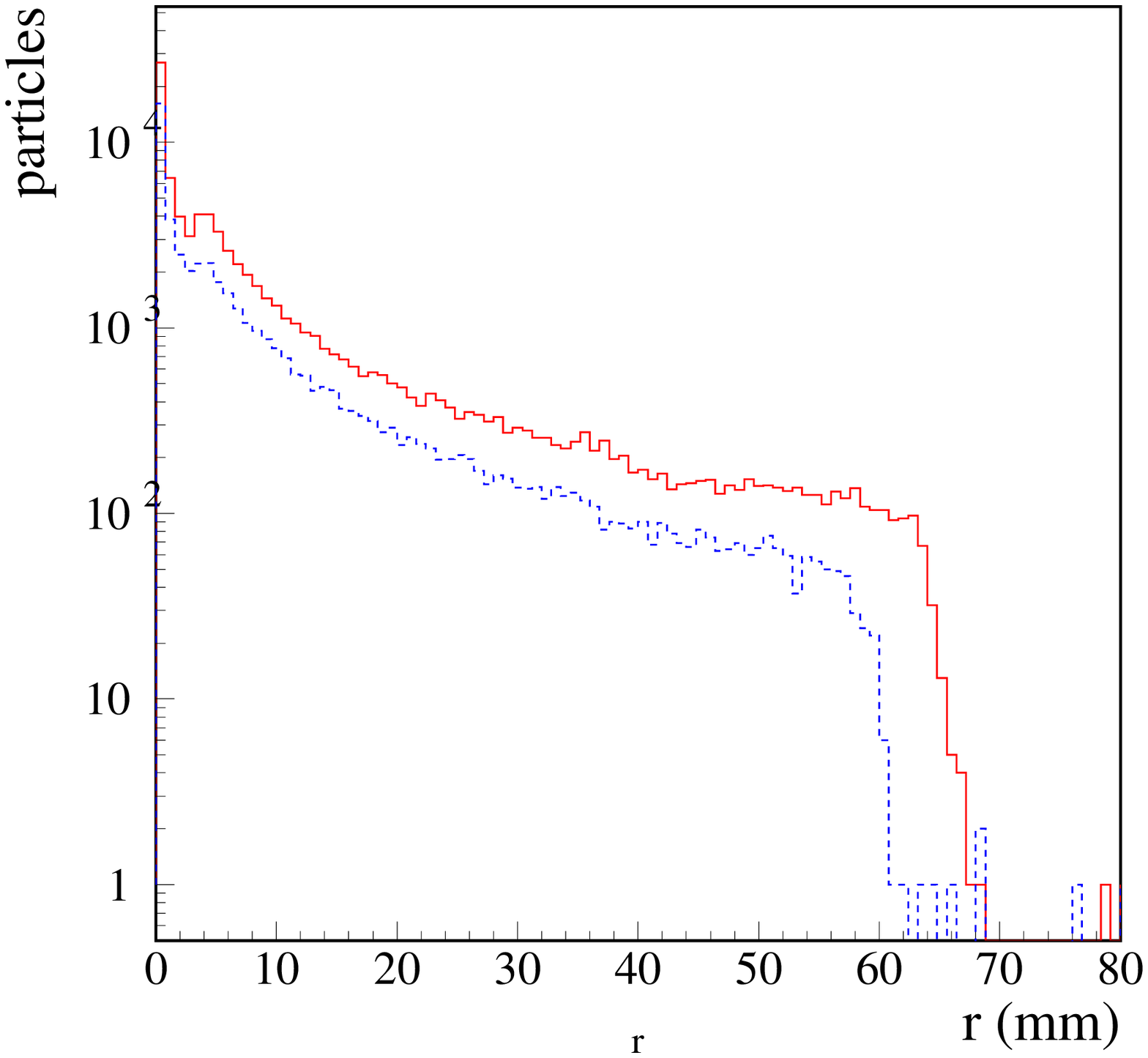}
\includegraphics[scale=0.32]{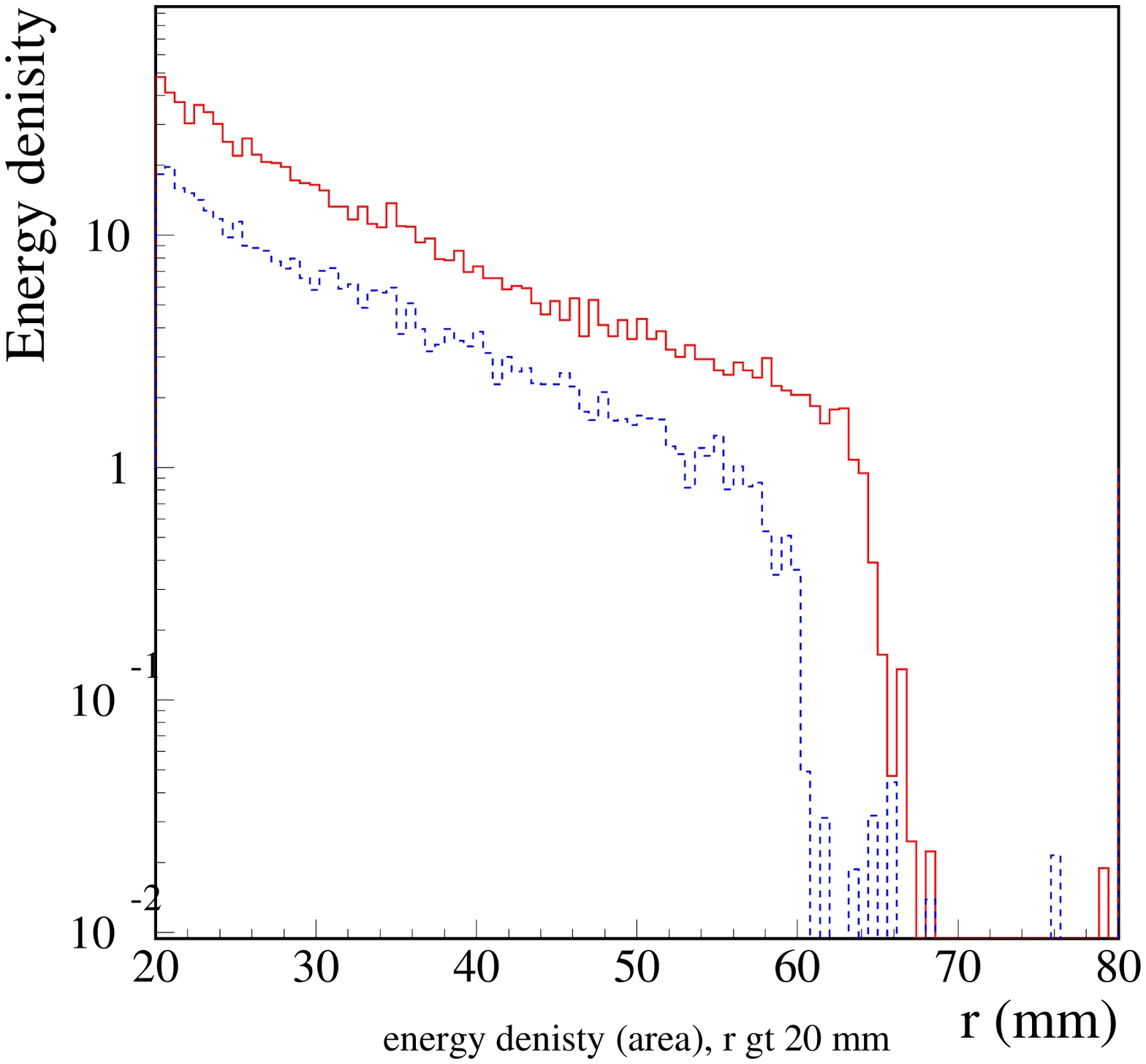}}
\caption{Hits (left) and energy-density (right) in
the ILD BeamCal. Solid red: SB2009, Dashed blue: RDR.}\label{Fig:bcalnumberanddensity}
%\end{wrapfigure}
\end{figure}

\section{SB2009 and physics}

Two examples will be used to show the impact on the performance
of the different designs: 
The analysis of the $\stau$ in the SUSY 
benchmark point SPS1a'~\cite{sps1ap}, which is sensitive to background
and polarisation,
and the measurement of the mass of a light SM Higgs with the recoil mass
method~\cite{higgsrecoil},
which is sensitive to the luminosity at lower $E_{CMS}$.

These two examples are not chosen at random. They both represent
corner-stone physics cases for the ILC in their own right.
This is even more so in view of the LHC program up to
2012: On one hand, the (probable) non-observation of the
Higgs in first LHC run would further strengthen the 
hypothesis that the Higgs is indeed SM-like and has a mass
close to the LEP exclusion limit. 
On the other hand, the current best fit of SUSY as an explanation
to all lower energy and cosmological observations, both those
that show no tension between the SM and the observations, and
those that do (eg. the muon g-2 or the cosmological evidence
for the existence of dark matter) is quite close to SPS1a'~\cite{ewfit}.
If this type of model turns out to be realised in nature,
the SUSY spectrum, including the squarks and the gluino, would
be light enough that there is a good chance that LHC will
discover SUSY in the first run~\cite{LHC}.
Hence, and due to the time it would require to modify accelerator
designs and to carry out a complete physics analysis,
it is prudent that the ILC community already now is
prepared to demonstrate that the ILC is right machine
for the future,
would any of these observations be done by the time of the
2012 summer conferences.
\subsection{SB2009 and physics: $\stau$ in SPS1a'}
The ILD analysis of the  $\stau$ in SPS1a' is described
in detail elsewhere~\cite{LOI}\cite{staupaper}\cite{peterthesis},
and only a few remarks are added here.

SPS1a' is a a pure mSUGRA model that 
predicts SUSY particles just outside what is 
excluded by LEP and low-energy
observations. 
It is compatible with the observations of WMAP, with the 
lightest SUSY particle as the Dark Matter candidate.
The LSP is the lightest neutralino, $\XN{1}$.
At $E_{CMS}=500 ~\GeV$, all sleptons are observable, but none of
the 
squarks.
The lighter bosinos, up to $\XN{3}$ (in  $\eeto \XN{1} \XN{3}$)
would also be observable. There are a total of 13 distinct channels
that would be observable below $E_{CMS}$=500 \GeV, a fact that demonstrates
the great advantage of a machine that can deliver high luminosities
at a large range of centre-of-mass energies. 

In SPS1a', the $\stone$ is the Next to Lightest SUSY Particle, the  NLSP,
and has a mass only 10 \GeV ~above the LSP mass ( 
$\mstone = 107.9~\GeV ,  \msttwo = 194.9~\GeV , 
\MXN{1} = 97.7~\GeVcc$).
The fact that the $\stau$ is the NLSP will imply that $\tau$:s 
are expected in most SUSY decays, so that SUSY will a main 
background to SUSY.
Finally, it can be noted that for 100 \% right positron polarisation, and
100 \% left electron polarisation ($\mathcal{P}_{beam}=(1,-1)$), the cross-sections
for $\XN{2} \XN{2}$ and $\XP{1} \XM{1}$ production is 
several hundred fb and  BR(X$ \rightarrow \stau) >$ 50 \%,
while 
for $\mathcal{P}_{beam}=(-1,1)$, these cross-sections almost vanish.
Therefore, the degree of polarisation of both beams is a very powerful
tool to ameliorate the signal to background ratio.

\begin{wrapfigure}{r}{0.5\columnwidth}
%\begin{figure}
\centerline{\includegraphics[scale=0.27]{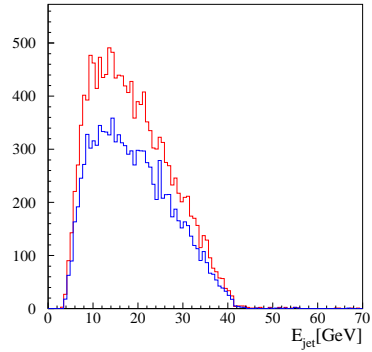}}
\caption{Spectrum of found $\tau$-jets. Blue: Durham, Red: DELPHI}
\label{Fig:taujetfinding}
\end{wrapfigure}
%\end{figure}
The mass of the $\stau$ can be obtained either
from the decay kinematics or the cross-section.
From the kinematics of two body decays it follows that
$E_{\tau, \genfrac{}{}{0pt}{}{min}{(max)}}\approx\frac{\sqrt{s}}{4}(1-(\frac{\MXN{1}}{\mstau})^2)
(1 \genfrac{}{}{0pt}{}{-}{(+)} \sqrt{1-\frac{4\mstau^2}{s}})$
(neglecting the $\tau$ mass).
For $\stone$ decays the resulting spectrum of the $\tau$:s has
$E_{\tau,min} = 2.6~\GeV , E_{\tau,max} = 42.5~\GeV$,
hence much of the signal will be hidden in the 
$\gamma \gamma$
background. 
For $\sttwo$ one finds that
$E_{\tau,min} = 35.0~\GeV , E_{\tau,max} = 152.2~\GeV$.
For such a spectrum, the background from
$W W \rightarrow l\nu l\nu$ will be an issue, while
the $\gamma\gamma$ background is expected to be small.
To determine the \stau ~mass 
from the spectrum, one needs to accurately
measure the end-point of the spectrum of the $\tau$ decay
products, which is equal to 
$E_{\tau,max}$.
In principle, the $\tau$ decay spectrum has a kink
at  $E_{\tau,min}$, but - at least for $\stone$ - this
region is completely hidden in the  $\gamma\gamma$ background.
Therefore, $\MXN{1}$ must be found from other channels, to be able
to determine the $\stau$ mass from the measurement of the end-point.
The $\stau$ production cross-section is given by
$\sigma_{\stau} = A(\theta_{\stau},\mathcal{P}_{beam}) \times \beta^3/s$, so
that a measurement of this cross-section can be used to determine
the mass by
$\mstau = E_{beam} \sqrt{1-(\sigma s/A)^{2/3}}$, which does not
depend on $\MXN{1}$.

For the ILD LOI~\cite{LOI}, all SM processes were fully simulated, as was the
full SPS1a' model. The
beam-background was estimated by generating 1000 bunch-crossings 
with GuineaPig, simulating the detector response, and reconstructing 
these with the full ILD reconstruction procedure, to create a pool of
bunch-crossings. For each physics event,
one bunch-crossing from the pool was randomly chosen and overlaid
before the event was analysed.

The properties of $\stau$ events are:
\begin{itemize*}
\item Only two $\tau$:s in the final state.
\item Large missing energy and momentum.
\item High acollinearity, with little correlation to the energy of the
$\tau$ decay-products.
\item Central production.
\item No forward-backward asymmetry.
\end{itemize*}
A set of cuts were found to select such events,
and to suppress the $\gamma\gamma$ background
further cuts were applied, among which was the requirement that
there should be no significant activity in the BeamCal,
see~\cite{staupaper}
for details. 
It should be noted that to get an acceptably low
background, it is paramount to have good $\tau$:s only, ie.
to have no extra or missing charged tracks.
In particular in the presence of beam-background, general jet-finders
perform poorly when used to find $\tau$:s.
Therefore the DELPHI $\tau$-finder~\cite{delphimssm} was adopted.
It was found to perform better than the Durham algorithm forced to two jets 
(the ILD default) already without
background, see Figure~\ref{Fig:taujetfinding}.

\begin{figure}
\centerline{\includegraphics[scale=0.3]{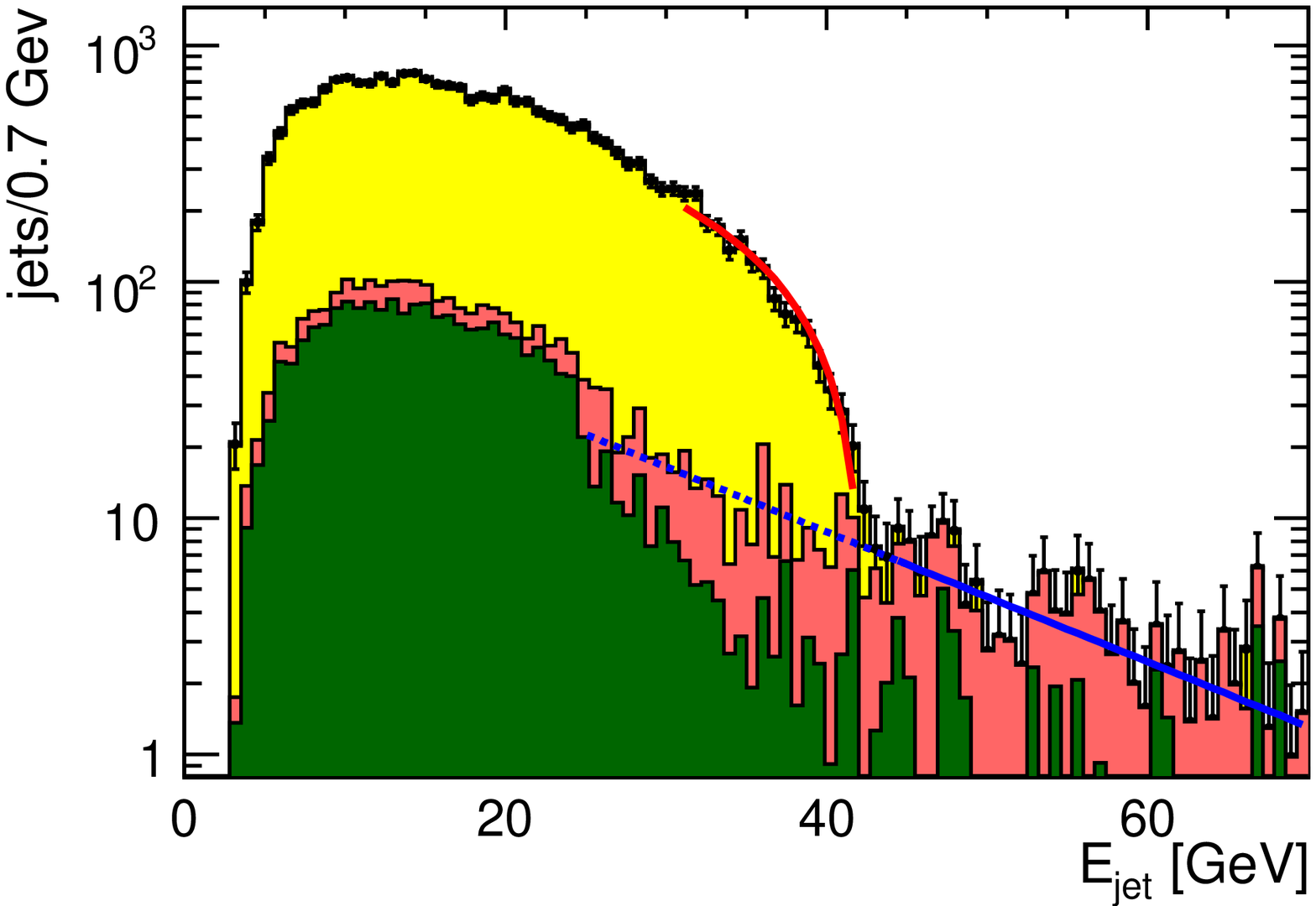}
\includegraphics[scale=0.3]{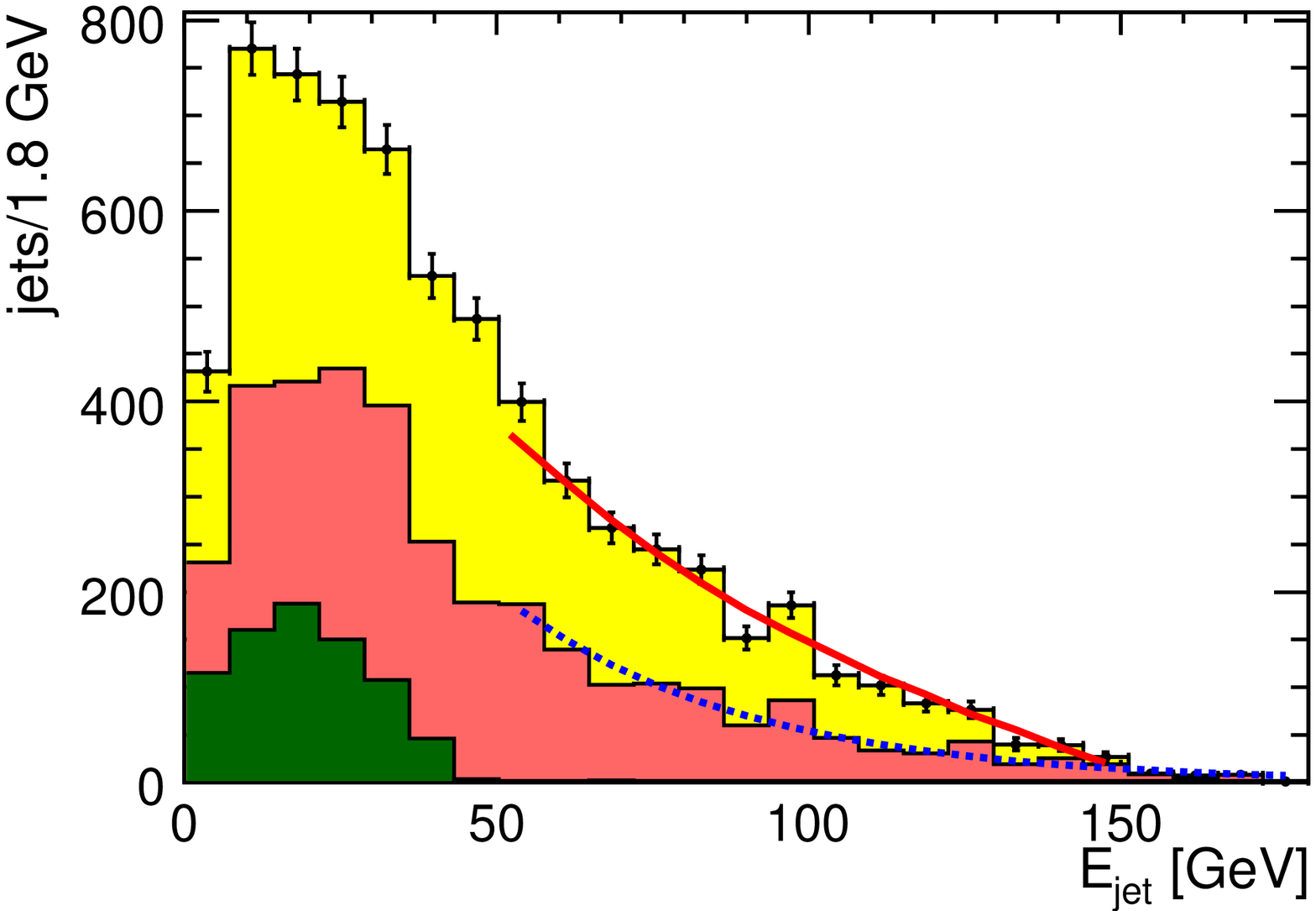}}
\caption{End-point fit for \stone (left) and \sttwo (right).
Light grey (yellow) histogram: signal, grey (red) : SM background, 
dark grey (green): SUSY background. The fit to
the background in blue. 
Fit to total sample: Solid (red) line.}\label{Fig:endpointfit}
%\end{wrapfigure}
\end{figure}

As has been emphasised above, only the upper end-point is relevant
for the determination of the mass from the spectrum.
The remaining background in this region needs to
be subtracted from the observed distribution,
and due to the difference in the amount of poorly known
SUSY background, this is done differently for $\stone$ and $\sttwo$.
In the case of \stone, a substantial amount of SUSY background remain
near the end-point,
which can be estimated from the data, by the
observation that the region above 45 $\GeV$ is signal-free. 
An exponential distribution was fitted to the data in this
region, and was extrapolated to lower jet energies.
For \sttwo, on the contrary, there is hardly any SUSY processes other than the signal
that gives jets
above 45 $\GeV$, so the background
can be estimated from SM-only simulation.
In both cases, the upper tail of the signal spectrum was
the obtained by fitting a line to the observed spectrum after
subtracting the fitted background. The spectra with signal and
background fits are shown in Figure~\ref{Fig:endpointfit}.
\begin{figure}[t]
\centerline{\includegraphics[scale=0.3]{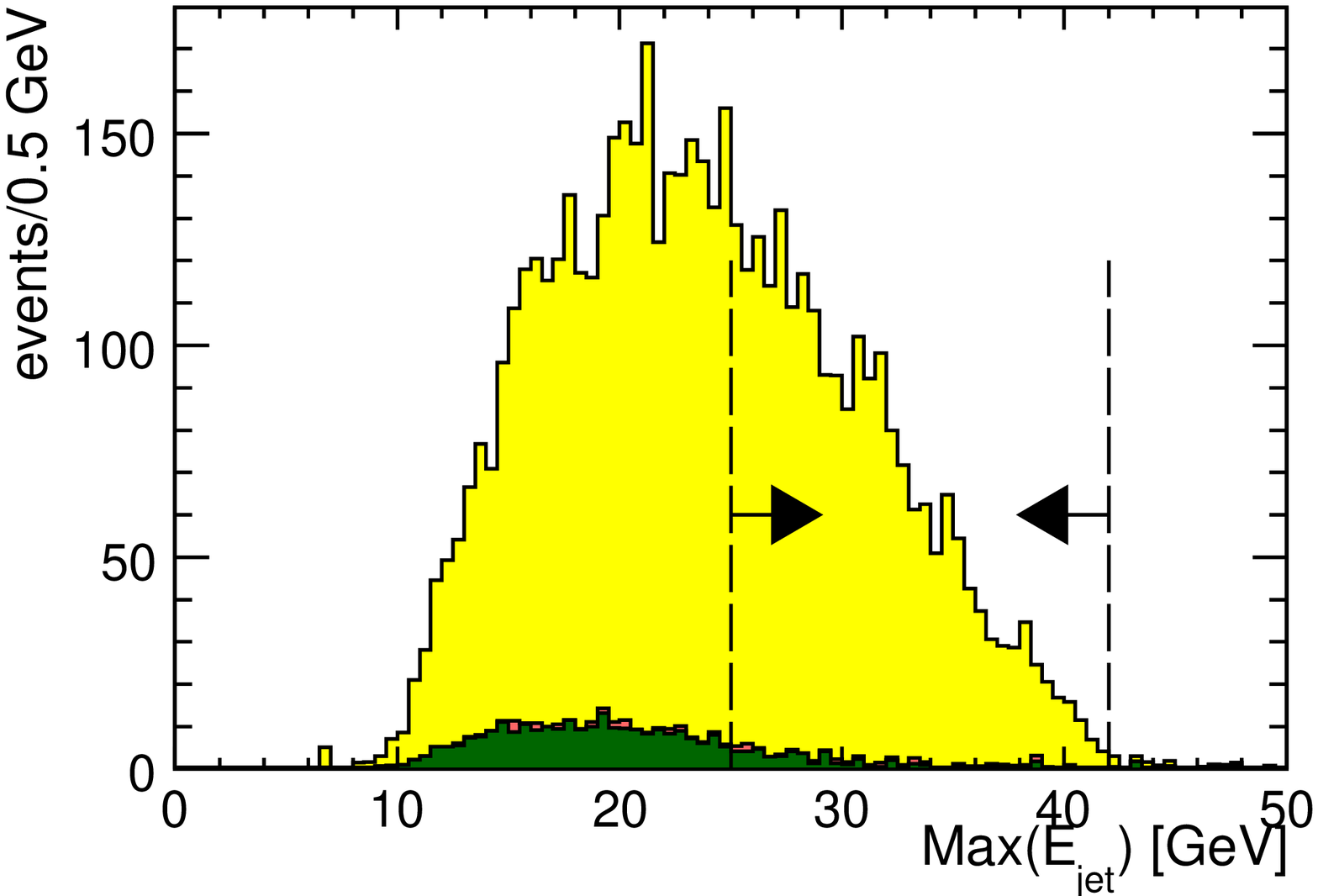}
\includegraphics[scale=0.3]{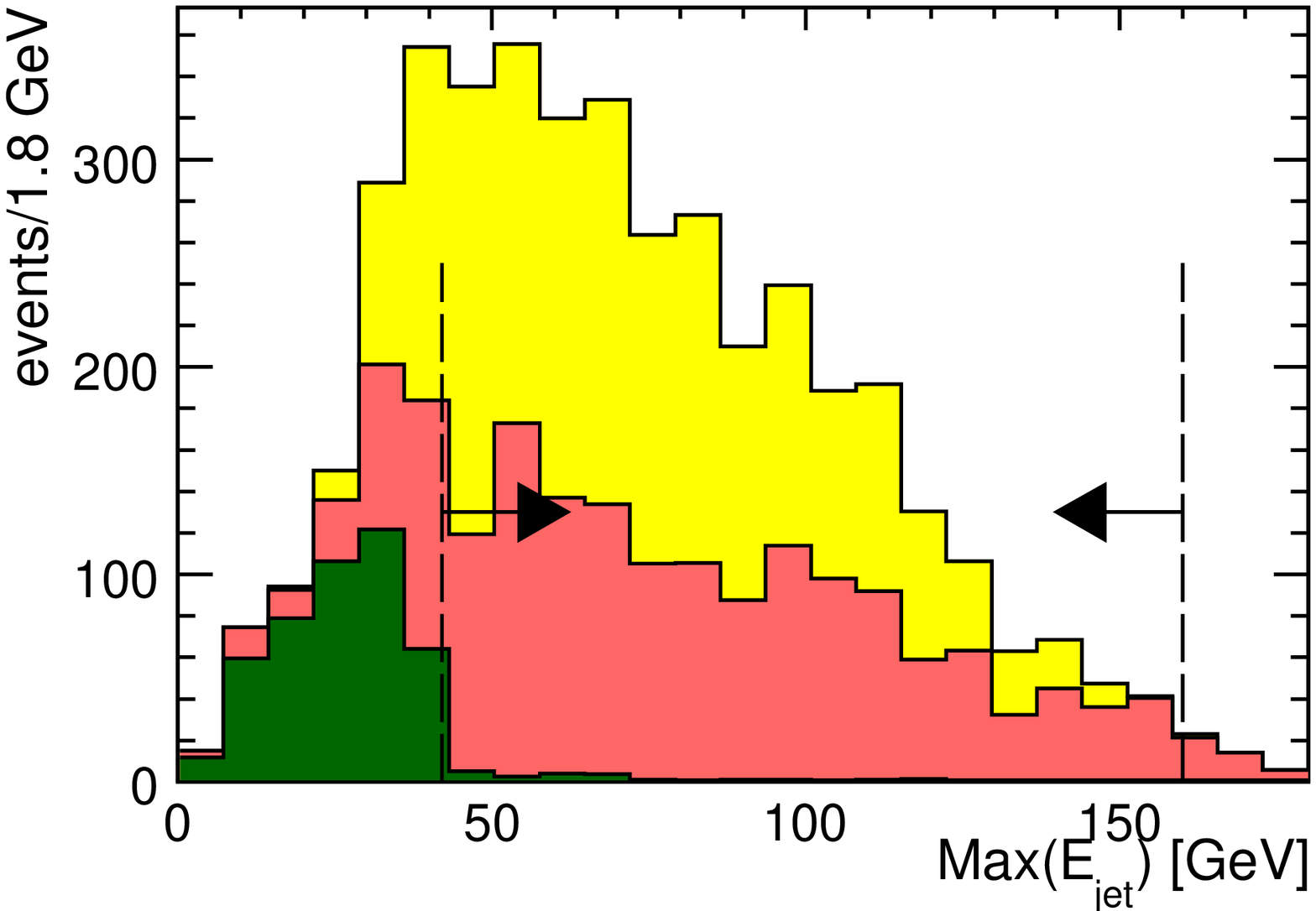}}
\caption{Events selected for the cross-section measurement for
\stone (left) and \sttwo (right). The selected events are
those with jet-energies in the indicated windows.
Legend as in Figure~\ref{Fig:endpointfit}.}\label{Fig:crosssection}
%\end{wrapfigure}
\end{figure}

The poorly known SUSY background is the most
important contribution to uncertainty of the cross-section
measurement, so efforts need to be made to
select region where is is as low as possible. The finally selected
events are shown in Figure~\ref{Fig:crosssection}.

The potential effects of the modified machine properties
for the $\stau$ analysis include:
\begin{itemize*}
\item The luminosity decrease for SB2009 without TF.
\item The decrease of $\mathcal{P}(e^+)$, which yields more SUSY background, 
and less signal for \stone.
\item The higher energy-spread of the incoming electron beam will blur the end-point.
\item The lower luminosity within 1 \% of nominal energy gives a lower signal
close to the end-point, where it has it's strongest significance.
\item The doubling the amount of beam-strahlung implies
more overlaid tracks (real or fake), which destroys the $\tau$ topology, and
twice as much energy in BeamCal, which increases the $\gamma\gamma$ background.
\item The higher luminosity in each bunch-crossing increases the probability
for a $\gamma\gamma$ event in the same bunch-crossing as the physics
event.
\end{itemize*}

\begin{table}[b]
%\begin{wraptable}{l}{0.5\columnwidth}

\centerline{\begin{tabular}{|l|c|c|c|c|c|c|}
\hline 
     & \multicolumn{6}{c|}{Events for end-point analysis} \\
\hline
case & \multicolumn{3}{c|}{\stone} &  \multicolumn{3}{c|}{\sttwo}\\
\hline
           &  SM    &     SUSY     &  signal &    SM    &    SUSY  & signal \\
\hline
 RDR       & 317    &     998      &   10466 &   1518   &    241   &  1983  \\
 SB09(TF)  & 814    &     956      &    8410 &   1346   &    223   &  1555  \\
 SB09(noTF) & 611    &     717      &    6308 &   1009   &    167   &  1166  \\
\hline
     & \multicolumn{6}{c|}{Events for cross-section analysis} \\
\hline
     & \multicolumn{3}{c|}{\stone} &  \multicolumn{3}{c|}{\sttwo}\\
\hline
           &  SM    &     SUSY     &  signal &    SM    &    SUSY  & signal \\
\hline
 RDR       & 17.6   &     47.7     &    2377 &   1362   &    33.7  &  1775  \\
 SB09(TF)  & 17.6   &     45.7     &    1784 &   1194   &    32.4  &  1366  \\
 SB09(noTF) & 13.2   &     34.3     &    1337 &    895   &    24.3  &  1025  \\
\hline
\end{tabular}}
\caption{Events after cuts in the \stau ~analysis for 
different designs (RDR, SB2009 with of without travelling focus).}
\label{tab:events}
%\end{wraptable}
\end{table}

The procedure to modify the fully simulated RDR sample
to represent the SB2009 design was as follows:
Our studies of the BeamCal indicates
that the energy density in either SB2009 design (with or without
TF) is about twice the RDR value at all radii.
For the LOI studies, the energy density from the beam-strahlung pairs
($\rho_E$) was mapped out over the surface of the BeamCal,
using the simulation 
with the RDR beam-parameters.
The probability $p(E_e , \rho_E)$
to detect an electron 
of energy $E_e$ in the BeamCal over a local 
energy density 
$\rho_E$ was then determined and parametrised.
Hence, the procedure factorises between $\rho_E$, which depends on
the beam-parameters, and $p(E_e , \rho_E)$, which doesn't.
Therefore, the SB2009 sensitivity for high-energy beam-remnants could be
estimated simply by scaling up $\rho_E$ from it's RDR value by a factor
determined to be 2.33,
and using the previously determined probability  $p(E_e , \rho_E)$.
To estimate how many tracks the beam-background would
create in the tracking system, a pool of fully simulated
bunch-crossings with SB2009 parameters was created and the
same procedure as for the LOI-RDR study (see above) was applied.
The different beam-spectrum was treated by using the
spectra obtained from GuineaPig for both RDR and
SB2009 to calculate event-by-event weights based on
the beam-energies of each individual event.
Finally, it was straight-forward to account for the 
reduced positron polarisation by correctly adjusting the
relative weights of  
samples 
generated with either $\mathcal{P}$=(-1,1) or $\mathcal{P}$=(1,-1).

\begin{table}
%\begin{wraptable}{l}{0.5\columnwidth}

\centerline{\begin{tabular}{|l|l|c|c|c|c|}
\hline
          &  &  \multicolumn{2}{c}{end-point (\GeV)}&
                \multicolumn{2}{|c|}{cross-section (\%)} \\
case      &\#&  \stone   &   \sttwo & \stone &    \sttwo\\
\hline
 RDR      & 1&  0.129    &    1.83  &  2.90  &     4.24\\
 +SB bck  & 2&  0.144    &    2.02  &  3.03  &     4.72\\
 +SB ppol & 3&  0.153    &    2.06  &  3.31  &     4.77\\
 +SB spect& 4&  0.152    &    2.10  &  3.52  &     5.09\\
 +SB noTF & 5&  0.179    &    2.42  &  3.79  &     5.71\\
\hline
\end{tabular}}
\caption{Errors on end-point and cross-section for the \stau ~analysis, with different beam conditions:
1) RDR, 2) RDR, but with background as with SB2009, 3) as 2) but with SB2009:s reduced
positron polarisation, 4) Nominal SB2009, with TF, ie. as 3) but with SB2009's beam-spectrum,
5) SB2009 without TF.}
\label{tab:errors}
%\end{wraptable}
\end{table}
\begin{figure}[t]
\centerline{\includegraphics[scale=0.3]{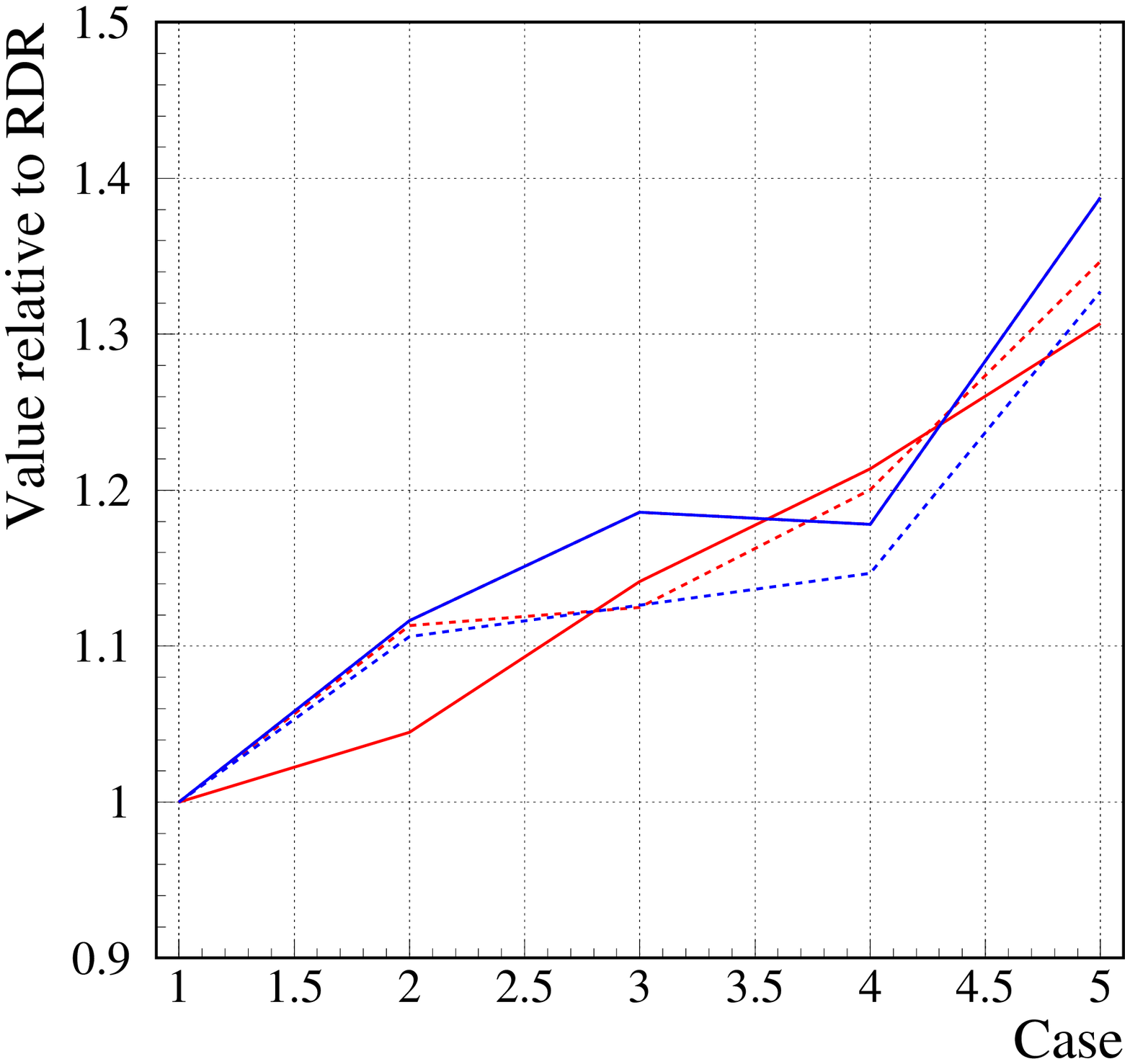}
\includegraphics[scale=0.3]{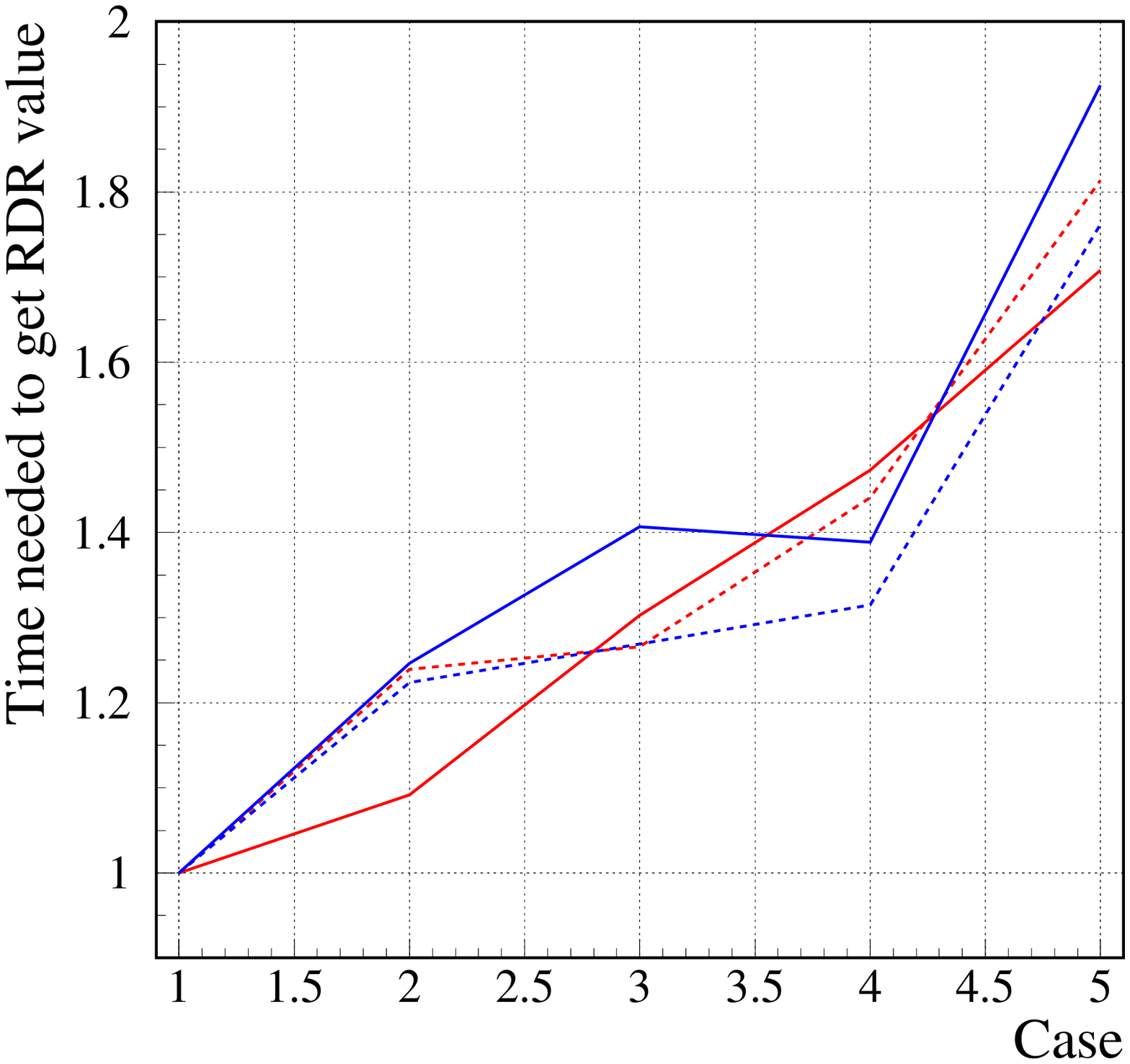}}
\caption {The effect of beam-parameters for the
results of the \stau ~analysis. The left plot shows
the uncertainty wrt. the RDR value, while the right one
shows the increase in data-taking time needed to achieve
the RDR value. Red: cross-section,
    Blue: end-point,
    Solid : \stone , Dashed: \sttwo.
The numbers on the x-axis are explained in caption of Table~\ref{tab:errors}. }\label{Fig:rdrtosbstau}
%\end{wrapfigure}
\end{figure}
Table~\ref{tab:events} shows the number of events after cuts,  
for the end-point and cross-section analyses, 
while Table~\ref{tab:errors} shows the error on the end-point  
and cross-section, for different machine
properties. In Figure~\ref{Fig:rdrtosbstau}, these numbers
are plotted both as the ratio of uncertainties with respect to the RDR value,
and as the increase in data-taking time needed to compensate
for the weaker performance.

%\small

\subsection{SB2009 and physics: SM Higgs at 120 \GeV}
The potential effect on the measurement of the mass of a 
light (120 \GeV) SM Higgs due to SB2009 modifications of
the ILC is the
factor 3 to 4 decrease of luminosity at optimal $E_{CMS}$($\approx$ 250 \GeV).
%LOI studies were done at 250 \GeV.
This reduced luminosity is, as has been explained
above, due to move of positron source.
Other aspects of SB2009 should pose no problems:
At $E_{CMS}=$ 250 \GeV, the 
undulator works at the same working-point for RDR and SB2009,
so no differences in incoming beam-spread nor positron polarisation
are expected.
The Higgs recoil-mass analysis, described in detail in~\cite{LOI} and~\cite{ILDhiggs}, 
only depends on the
precise measurement of the decay of the Z to high momentum muon or
electron pairs.
Hence, it is not sensitive to $\gamma\gamma$
background, nor to overlaid background tracks.

It has been suggested to do the analysis at 350 \GeV, where luminosity
loss is much 
less important(20-40 \%). 
However, the cross-section is sizably lower at this energy,
and since the lepton-pairs will have higher momenta,
the experimental resolution will be worse.
Since this channel was not studied
at 350 \GeV ~for the LOI,
a fast simulation was set up to bring the
experience gained from full simulation at 250 \GeV ~to
an estimate of the performance at 350~\GeV, both for the RDR and SB2009
designs~\cite{hengnefast}.
The fast simulation has been verified with respect to the full simulation at 250~\GeV, 
with RDR parameters, and  excellent agreement was found.

%\begin{wrapfigure}{r}{0.5\columnwidth}
\begin{figure}[b]
\centerline{\includegraphics[scale=0.3]{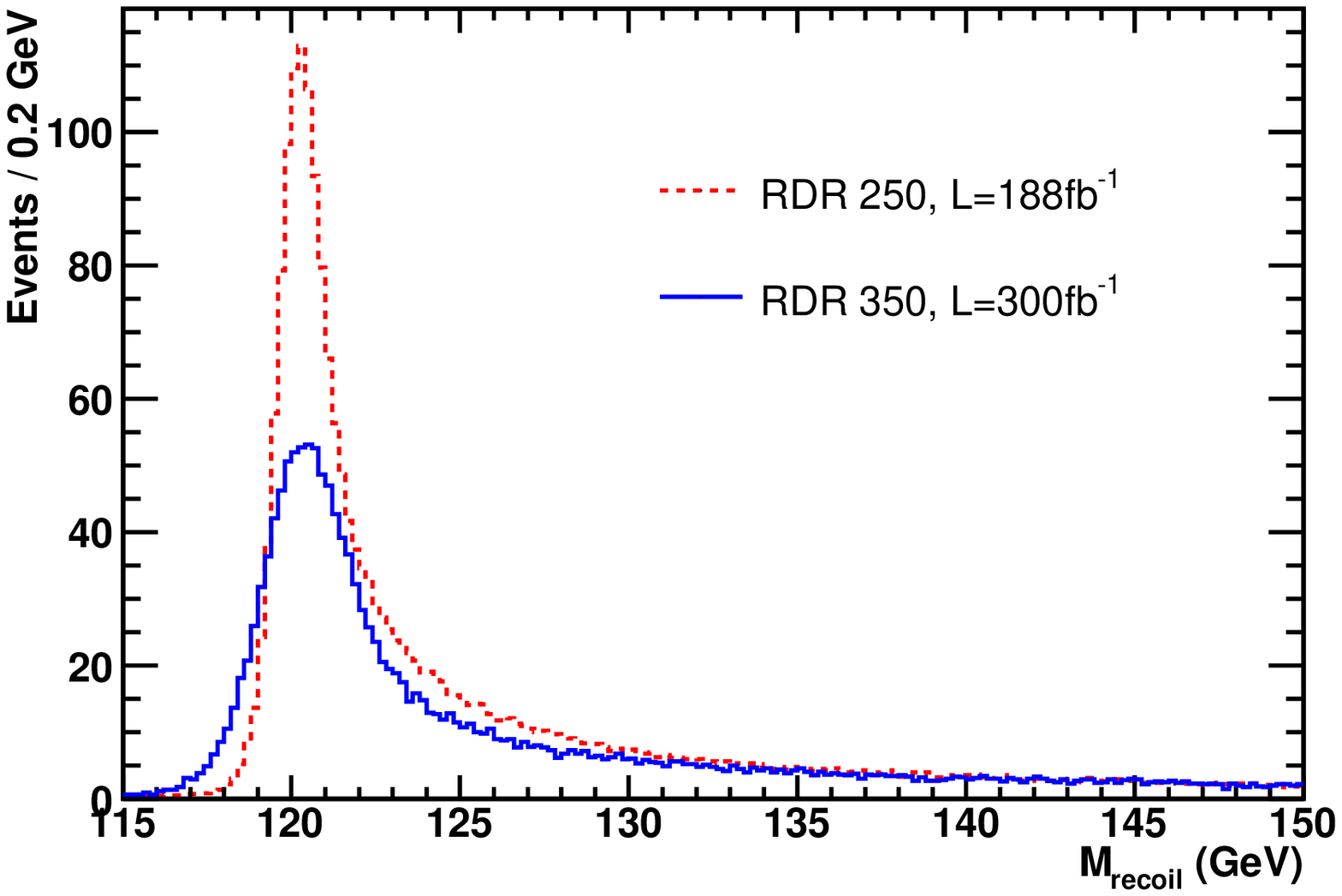}
\includegraphics[scale=0.3]{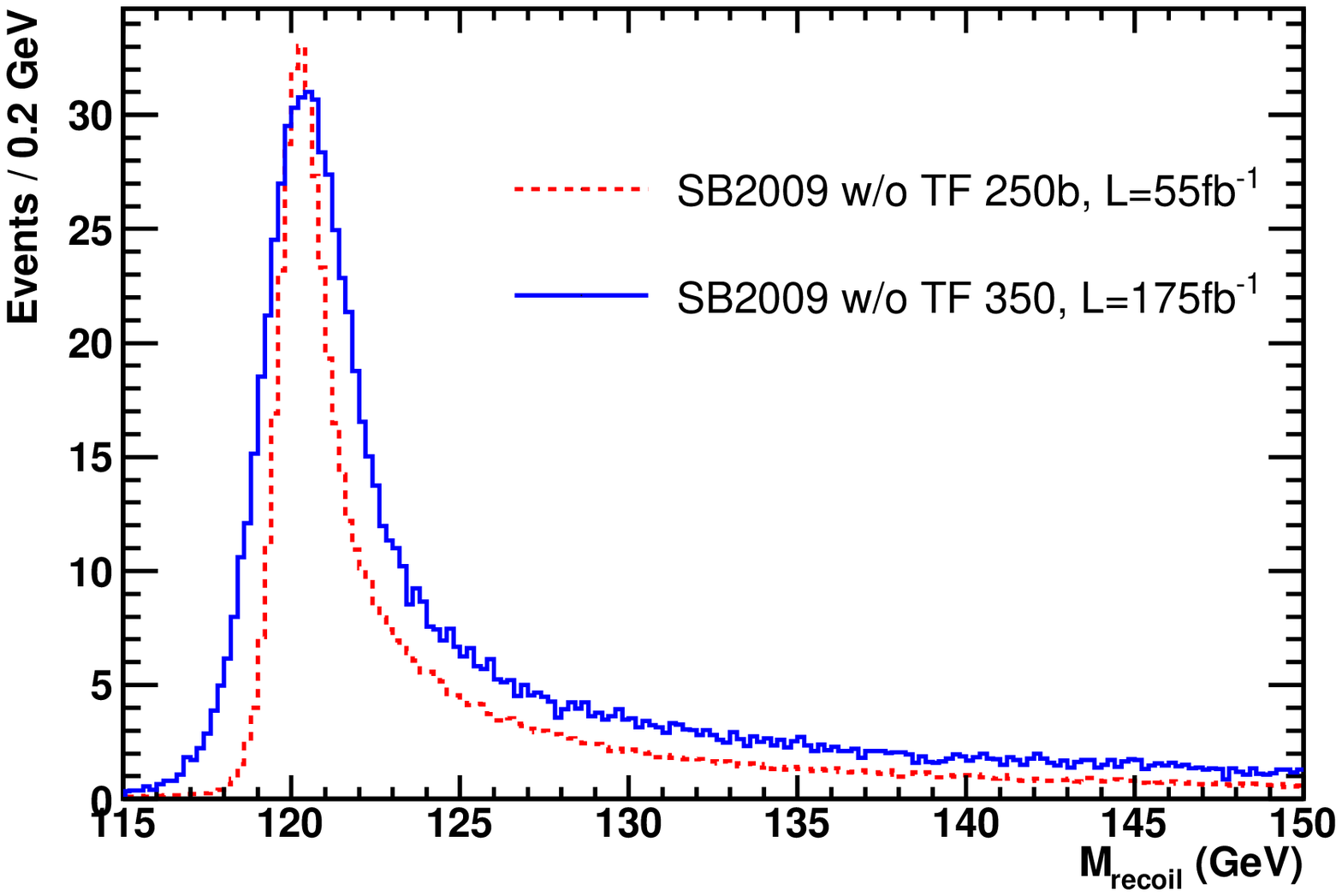}}
\caption{Recoil Mass for the RDR design (left) and the SB2009, without 
TF (right) for $E_{CMS}$=250 and 350 \GeV}\label{Fig:higgsrdrsn-250-350}
%\end{wrapfigure}
\end{figure}
One could then compare the recoil-mass peak obtainable with the
same running-time at 250 or 350 \GeV, for RDR and SB2009.
Two observations can be done in Figure~\ref{Fig:higgsrdrsn-250-350}:
As expected, there is a much larger loss of events going from
350 to 250 \GeV ~in the SB2009 design, than what is observed for the RDR
design.
Furthermore, for both designs, the peak is broader at 350 ~\GeV,
which, as can be seen in Figure~\ref{Fig:higgsgen-sim}, is due to the deterioration
of the momentum resolution at higher track moments.
Table~\ref{tab:higgserrors} shows the details of the deterioration of
the measurement wrt. the LOI design.

\begin{table}
%\begin{wraptable}{l}{0.5\columnwidth}

\centerline{    \begin{tabular}{|l|c|c|c|c|c|} 
    \hline
    Beam Par & $\mathcal{L_{\rm int}}$ (fb$^{-1}$) & $\epsilon$ & S/B & $\Delta(M_{H})$ (GeV)& $\delta\sigma/\sigma$ \\ 
    \hline
    RDR 250 &		188	& 55\% & 62\% & 0.043 & 3.9\% \\
    RDR 350 &		300	& 51\% & 92\% & 0.084 & 4.0\% \\
    SB2 TF 250b &	68	& 55\% & 62\% & 0.071 & 6.4\% \\
    SB  TF 350 &	250	& 51\% & 92\% & 0.092 & 4.3\% \\
    SB2 noTF 250b &	55	& 55\% & 62\% & 0.079 & 7.2\% \\
    SB Wolf 350 &	175	& 51\% & 92\% & 0.110 & 5.2\% \\
    \hline
    \end{tabular}}
\caption{Performance of the Higgs recoil mass analysis for different beam conditions.}
\label{tab:higgserrors}
%\end{wraptable}
\end{table} 

%\small

\begin{wrapfigure}{r}{0.5\columnwidth}
%\begin{figure}[b]
\centerline{\includegraphics[scale=0.3]{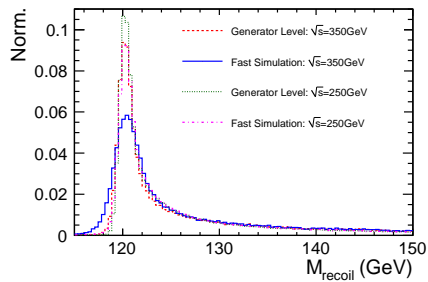}}
\caption{Generated and simulated Recoil Mass for the RDR design,
at $E_{CMS}$=250 and 350 \GeV}\label{Fig:higgsgen-sim}
\end{wrapfigure}
%\end{figure}
One can note that for the RDR design, the 250 \GeV~result 
is best both for cross-section and mass measurements,
while for the SB2009 (TF), the  250 \GeV~result is best for the mass 
measurement, while the 350 \GeV~is the best for the cross-section measurement.
Hence, for SB2009, one must choose which running scheme to use:
either the mass-measurement will worse by 110 \%, 
but the cross-section only by 10 \% with respect to the RDR design,
or the mass-measurement will be worse by a more modest 70 \%, 
at the expense that the cross-section will be worse by 65 \%.

\section{Conclusions and outlook}

In section 1, the ILC was briefly described, and it was 
pointed out that depending on what 
is built in to the machine design ($P_{RF}$, $f_{rep}$,
$N_{bunch}$, $\delta_{BS}$ ...), luminosity scales differently
with $E_{cm}$,
and that different machine setups give
different luminosity scaling, different polarisation scaling,
different energy within 1 \% to nominal, and different spread 
in $E_{beam}$.
The key issues for performance were pointed out, and
how they relate to technological challenges for the accelerator
design.

From studying the effect of the transition from
the RDR to the SB2008 designs on the \stau ~analysis it was seen
that for such ``fragile'' signals, beam-background influences signal 
directly.
It was also seen that for any ``low $\Delta(m)$'' ($< 10$ GeV) signal, 
beam-background should
be taken into 
consideration when estimating $\gamma\gamma$ background.
It was found that effect of replacing the RDR 
beam-parameters by the SB2009 ones, the measurement quality
degrades by 15-20 \%, both for
the end-point and cross-section measurements, and both for the 
$\stone$ and $\sttwo$ channels. 
Of this degradation, half comes from the modifications
of the positron source, namely the larger spread in $E_{CMS}$, 
and the reduction in $\mathcal{P}(e^+)$. 
It was also shown that the degradation increases to 20-40 \%,
if the travelling focus concept would turn out to be unfeasible.
It was pointed out that degradation can be compensated for
by increasing the data-taking period by 40 \% (or 80 \%
in the no travelling focus case).
It was also pointed out that for a low-mass SUSY scenario like SPS1a',
many thresholds are expected below $E_{CMS}$=500 \GeV,
emphasising the importance of having an accelerator tunable in
energy.

By studying the effects of the transition on the SM Higgs
recoil-mass analysis, it was observed
that 
the results will not scale with cross-section
if the analysis is done a different $E_{CMS}$, since the
detector resolution depends on energy. At $E_{CMS}$ = 250 \GeV,
the influence on the error on the Higgs mass from the detector
resolution is quite small compared to the spread of the
beam energy, but already at $E_{CMS}$ = 350 \GeV,
detector effects dominate the error.
It was found that  replacing the RDR design by the SB2009 one
degrades the quality of the mass measurement by 110\%, and the
cross-section measurement by 10 \% (if the analysis is done
at 350 \GeV), or by 70 \% and 65 \% (if it is done at 250 \GeV).
Without travelling focus, the numbers are 160 \% and 30 \% (at 350 \GeV)
or 80 \% and 85 \% (at 250 \GeV). To compensate
for these losses in precision by longer data-taking
is difficult: at least three times more time need
to be spent at an energy where the Higgs channel
is likely to be the only channel to study.
The loss in performance is in this case completely 
driven by loss of luminosity at 250 \GeV, which is entirely
due to move of positron source. 

Hence, both these studies arrive at the conclusion that the
move of the positron source has a disproportionally large effect
on the performance, while the increased background has
less of an impact.

Since the presentation of the SB2009 proposal in December 2009,
it has been reviewed by the AAP~\cite{aap} and the PAC~\cite{pac}, 
and a committee
(the ``Brau committee'') has been set up to review the performance
issues the proposal gave rise to. In particular, the effect
of the move of the positron source has received much attention.
The GDE has subsequently set up a series of work-shops aiming to solve
this and other issues by mid-2011, and defined a formal 
``change control process'' for the baseline design of the ILC.

There are no significant financial savings from placing
the positron source at the end of the linac. 
However, from the 
operational point of view,
it is strongly preferred to have it at that location, 
since this would  keep the main linac tunnel free
of a potentially delicate sub-system,
and rather concentrate it with all other sub-systems (the electron source,
the damping rings, BDS, etc.)
to a central campus, where access and maintenance would be eased.
Several promising schemes are investigated
to avoid the large loss of luminosity below $E_{CMS}=$300 \GeV,
while still placing the source at the end of the linac.
The most probable solution is to use the fact that at lower
beam-energies, there is enough spare RF-power to
increase the repetition rate to 10 Hz. One would then accelerate  every other
electron bunch-train to 150 \GeV ~to produce positrons,
and every other, at a lower energy, would be sent to the
detector.
The feasibility of this scheme is under study, in particular
with respect to which modifications would be needed to the
damping system to accommodate a higher repetition rate.

%\section{Bibliography}

%If possible please use the bibtex information as given by SPIRES
%to make the citations~\cite{parton_qed} uniform and follow the 
%examples~\cite{parton_qed,H1,DVCS,pomeron} given below.
%Note that there is a (non-breaking) space before \verb?\cite?.

% ****************************************************************************
% BIBLIOGRAPHY AREA
% ****************************************************************************

\begin{footnotesize}
% IF YOU DO NOT USE BIBTEX, USE THE FOLLOWING SAMPLE SCHEME FOR THE REFERENCES
% ----------------------------------------------------------------------------

% ----------------------------------------------------------------------------

\end{footnotesize}

% ****************************************************************************
% END OF BIBLIOGRAPHY AREA
% ****************************************************************************

\end{document}

%% file: newcom.tex
\def\leqsim{\mathbin{\;\raise1pt\hbox{$<$}\kern-8pt\lower3pt\hbox{$\sim$}\;}}
\def\geqsim{\mathbin{\;\raise1pt\hbox{$>$}\kern-8pt\lower3pt\hbox{$\sim$}\;}}
\def\MXN#1{\mbox{$ M_{\tilde{\chi}^0_#1}                                $}}

\def\XP#1{\mbox{$ \tilde{\chi}^+_#1                                     $}}
\def\XM#1{\mbox{$ \tilde{\chi}^-_#1                                     $}}

\def\XN#1{\mbox{$ \tilde{\chi}^0_#1                                     $}}

\def\p#1{\mbox{$ \mbox{\bf p}_1                                         $}}

\newcommand{\stau}    {\mbox{$ \tilde{\tau}                                $}}
\newcommand{\stone}   {\mbox{$ \tilde{\tau}_1                              $}}
\newcommand{\sttwo}   {\mbox{$ \tilde{\tau}_2                              $}}

\newcommand{\mstau}   {\mbox{$ M_{\tilde{\tau}}                            $}}
\newcommand{\mstone}  {\mbox{$ M_{\tilde{\tau}_1}                          $}}
\newcommand{\msttwo}  {\mbox{$ M_{\tilde{\tau}_2}                          $}}

\newcommand{\eeto}    {\mbox{$ {\, \mathrm e}^+ {\mathrm e}^- \to             $}}

\newcommand{\MeV}     {\mbox{$ {\mathrm{MeV}}                              $}}

\newcommand{\GeV}     {\mbox{$ {\mathrm{GeV}}                              $}}

\newcommand{\GeVcc}   {\mbox{$ {\mathrm{GeV}}/c^2                          $}}

\newcommand{\ba}{\begin{array}}
\newcommand{\ea}{\end{array}}
\newcommand{\bc}{\begin{center}}
\newcommand{\ec}{\end{center}}
\newcommand{\be}{\begin{eqnarray}}
\newcommand{\eeq}{\end{eqnarray}}
\newcommand{\bes}{\begin{eqnarray*}}
\newcommand{\ees}{\end{eqnarray*}}
\newcommand{\Kz}{\ifmmode {\rm K^0_s} \else ${\rm K^0_s} $ \fi}
\newcommand{\Zz}{\ifmmode {\rm Z^0} \else ${\rm Z^0 } $ \fi}
\newcommand{\xxbar}{\ifmmode {\rm x\bar{x}} \else ${\rm x\bar{x}} $ \fi}
\newcommand{\rphi}{\ifmmode {\rm R\phi} \else ${\rm R\phi} $ \fi}
\def    \missEt      {\ifmmode{/\mkern-11mu E_t}\else{${/\mkern-11mu E_t}$}\fi}
\def    \missE       {\ifmmode{/\mkern-11mu E}\else{${/\mkern-11mu E}$}\fi}
\def    \missp       {\ifmmode{/\mkern-11mu p}\else{${/\mkern-11mu p}$}\fi}
\def    \misspt      {\ifmmode{/\mkern-11mu p_t}\else{${/\mkern-11mu p_t}$}\fi}